\documentclass[aps,twocolumn,preprint,color,psfig,epsf]{revtex4}
\usepackage{amsmath}
\usepackage{amsfonts}
\usepackage[dvips]{graphicx}
\begin{document} 
\title{N-dimensional electron in a spherical potential: the large-N limit}
\author{Amit K Chattopadhyay}
\address{Dipartimento di Fisica ``Galileo Galilei'', Universita' di Padova,
via Marzolo 8, 35131 Padova, Italy}
\email{amit@pd.infn.it} 
\date{\today} 
\begin{abstract} 
\noindent 
We show that the energy levels predicted by a $1/N$-expansion method for an N-dimensional
Hydrogen atom in a spherical potential are always lower than the exact energy levels but
monotonically converge towards their exact eigenstates for higher ordered corrections. The
technique allows a systematic approach for quantum many body problems in a confined potential
and explains the remarkable agreement of such approximate theories when compared with 
the exact numerical spectrum. \\

\noindent
Keywords: $1/N$-expansion, hyperspherical coordinates. \\

\noindent
PACS: 03.65.-w, 73.21.La
\end{abstract} 
 

\maketitle 

\newpage

A fundamental theoretical problem in the realm of many body physics concerns 
the technical difficulty in making precise theoretical evaluations of physical
observations, even more so in problems involving quantum systems. In most cases, 
this is compounded by the practical difficulty in
establishing a suitable approximation scheme that conjoins simplicity with 
effectiveness. An interesting observation in connection suggests 
that {\em an increase in the number of degrees of freedom often simplifies 
the theoretical analysis} \cite{witten}. A perturbative approach requires at least one
dimensionless parameter and if we couple this fact with the previous statement, 
it effectively implies that as we 
go on increasing the dimensionality of this parameter, the perturbation analysis 
becomes more and more simple \cite{wilson,hooft}. Often it is found that a problem of
inherently quantum mechanical origin can be mapped on to a classical phase space in  
the $N \rightarrow \infty$ limit thereby reducing a quantum problem to a classical one
\cite{yaffe}. In other words, one then has a limit where quantum interference effects
simply die out paving the way for a simple classical analysis. The excited states 
for such a system can be obtained as an expansion
in $1/N$ around the minimum of the effective classical potential $V_{\mathrm{eff}}$. Such
an approach is not at all uncommon in statistical physics \cite{Berlin_Kac,spin} in problems
which allow for at least a minimum. In many of those cases, the large-N limit has 
been fruitfully utilized in dealing with equilibrium as well as non-equilibrium problems 
in classical critical phenomena \cite{Ma}. In quantum mechanics too, $1/N$ expansion
method has a long precedence. Detailed accounts of related applications can be obtained 
from review articles like the one due to Chatterjee \cite{moshe,aharony,chatterjee,cohen} 
as also from more informal narratives like \cite{witten,yaffe}. The versatility and flexibility
of this technique has allowed it to be used in a range of diverse topics, starting
from field theoretic studies in high energy physics \cite{gravity,yang_mills} to 
problems on earthquake dynamics \cite{earthquake} as well as on problems in colloidal 
physics \cite{colloids}.

In this brief report, we shall use the $1/N$-expansion method to study the problem of an 
N-dimensional Hydrogen atom confined in a Harmonic oscillator potential. Although the 
model is nothing new \cite{chatterjee}, however our objective here is. We intend to 
study the efficacy
of this expansion method by calculating the energy eigenvalues and showing that to each
order of correction, the large-N expansion method always predicts a slightly lower potential
as compared to the exact eigenvalue obtained numerically. This is remarkable since this implies
a certain monotonicity in these perturbation corrections which tells us that the corrections
are always {\em positive}, a fact that has often been {\em tacitly assumed} in related 
calculations \cite{el-said,garcia-castellan}. We argue that this is the underlying reason
which makes this method more dynamic compared to standard perturbation technique which
is limited strictly to a {\em weak-coupling} regime. In a following work, we build
on this principle and analytically solve for the three-body problem of interacting
electrons using an {\em exact} Coulomb potential \cite{chattopadhyay}. 

In the first paper of the paper, we do a rehash of the N-dimensional quantum
mechanics for a single electron in a spherical confining potential and then 
defining the potential in the relative frame of reference, we go on 
to solve the stationary state problem using the $1/N$-expansion method. As already 
stated, we then
proceed to calculate the energy corrections due to this method for both ground and excited
states and show that all higher ordered corrections have a steady monotonicity that ensures
a large-N eigenstate below its exact (meaning experimental) counterpart. 

Taking cues from standard literature \cite{chatterjee,moshe}, we begin with the 
Hamiltonian for the center of mass of an N-dimensional electron in a spherical potential 

\begin{equation}
H = \frac{{\vec p}^2}{2m_e} + V_N(\vec r)
\end{equation}

Using standardized units ${\bar h}=m_e=1$ (h=Plank's constant and 
$m_e$=mass of the electron), the Hamiltonian can be rewritten as 

\begin{equation}
H = -\frac{1}{2} {\nabla_N}^2 + V_N(\vec r)
\end{equation}

where terms have their usual meaning. The potential being radial
$V_N(\vec r)=V_N(r)$ and this gives the eigenvalue equation 

\begin{equation}
H \psi(\vec r) =
[-\frac{1}{2} {\nabla_N}^2 + V_N(\vec r)] \psi(\vec r) = E \psi(\vec r)
\label{equation}
\end{equation}

For a system with spherical symmetry, the curvilinear coordinates can be written as follows
(generalisation of the treatment available in \cite{arfken})

\begin{eqnarray}
x_1 &=& r \cos{\theta_1} \sin{\theta_2} \sin{\theta_3} ... \sin{\theta_{N-1}} \nonumber \\
x_2 &=& r \sin{\theta_1} \sin{\theta_2} \sin{\theta_3} ... \sin{\theta_{N-1}} \nonumber \\
x_3 &=& r \sin{\theta_2} \sin{\theta_3} \sin{\theta_4} ... \sin{\theta_{N-1}} \nonumber \\
. \nonumber \\
. \nonumber \\
. \nonumber \\
x_k &=& r \cos{\theta{k-1}} \sin{\theta_k} \sin{\theta_{k+1}} ... \sin{\theta_{N-1}} 
\nonumber \\
. \nonumber \\
. \nonumber \\
. \nonumber \\
x_{N-1} &=& r \cos{\theta_{N-2}} \sin{\theta_{N-1}} \nonumber \\
x_N &=& r \cos{\theta_{N-1}}
\label{spherical_coordinates}
\end{eqnarray}

where $r$ is the r is the radial distance and $\theta_k (k<N-1)$ are the angles defining the 
hyper-spherical space, $\theta_{N-1}$ being the azimuthal angle. $\psi(\vec r)$ is the
eigenfunction of this system. The above definition can
now be used to obtain the radial equation of motion \cite{chatterjee}

\begin{eqnarray}
&[& -\frac{1}{2}(\frac{d^2}{dr^2} + \frac{N-1}{r} \frac{d}{dr}) + \frac{l(l+N-2)}{2r^2} 
\nonumber \\
&+& V_N(r)] R(r) = E R(r)
\label{basic_radial_equation}
\end{eqnarray}

where $l$'s are the angular quantum numbers and $R(r)$ is the radial wave function. 
Using the transformation $u(r)=r^{(N-1)/2} R(r)$, we can now absorb the first derivative
in eqn. (\ref{radial_equation}). The reconstructed radial equation of motion is now given by

\begin{equation}
-\frac{1}{2} \frac{d^2 R}{dr^2} + k^2 [\frac{(1-\frac{1}{k})(1-\frac{3}{k})}{8r^2} 
+ \frac{V_N(r)}{k^2}] u(r) = E u(r)
\label{radial_equation}
\end{equation}

In the above, we have used $k=N+2l$. At this point, the meaning of the large-N limit turns
out to be pretty obvious. It means that $k \rightarrow \infty$ (since N is large)
encompasses the idea of a stationarity limit for a very heavy {\em classical} 
particle of effective mass $k^2$ where the particle is localized at the point $r=r_0$, 
the point $r_0$ in turn defining the minimum of the classical
potential $V_{\mathrm{eff}}=\frac{1}{8r^2} + \frac{V_N(r)}{k^2}$. The ground state energy
of such a localized system is given by $E_{\infty}=k^2 V_{\mathrm{eff}}(r_0)$. 

We now consider a specific form for the potential function $V_N(r)$ and proceed to
calculate the higher order corrections in the large-N limit. The model we choose
for the purpose is an oscillator with anharmonic fluctuations. The reason for this
choice has been accentuated by the observation that such a description, albeit simple,
yet is able to reproduce a good estimate for the energy eigenstates \cite{johnson_payne1991}
when compared with numerical \cite{chakraborty} as well as with experimental \cite{experiment}
result. For a simple harmonic oscillator $V_N(r)=\frac{1}{2} \omega^2 r^2$ which gives
$r_0=\sqrt{\frac{k}{2\omega}}$, $V_{\mathrm{eff}}=|\frac{\omega}{2k}|$ and eventually
$E_{\infty}=|\frac{3\omega}{2}|$. One can now add quantum fluctuations and study the
behavior of the system close to the classical minimum $r_0$ \cite{chatterjee}. We go beyond 
this description in the sense that we consider a {\em finite sized} electron instead
of a fixed mass and consider fluctuations around the classical stable minimum. To do this
we revoke the original radial equation eq. (\ref{radial_equation}) prior to the large-N
limit being imposed on it. Using the $1/N$ expansion technique, we now embark on a 
stepwise evaluation of the energy eigenvalues due to the quantum fluctuations close 
to the classical minimum. We define the eigenvalue problem as follows

\begin{equation}
[H_0 + {\hat V(r)}] \psi(\vec r) = E \psi(\vec r)
\label{eigenvalue_equation}
\end{equation}

The ground state eigenvalue equation $H_0 \psi(\vec r) = E_0 \psi(\vec r)$ has already
been defined through equation (\ref{equation}) ($E_0=\frac{3\omega}{2}$) 
while ${\hat V(r)}$ is the part of the Hamiltonian that contributes to the quantum fluctuations.
Taylor's expansion allows this perturbation Hamiltonian to be represented as 

\begin{equation}
{\hat V} = {\hat V}(r_0) + (r-r_0) {\hat V}'(r_0) + \frac{{(r-r_0)}^2}{2!} {\hat V}''(r_0) + ...
\end{equation}

where the primes denote derivatives with respect to r.
Before proceeding any urther, we make a variable transformation from $r \rightarrow x$ where
$x=\frac{\sqrt{k}}{r_0}(r-r_0)$ and transform eq. (\ref{eigenvalue_equation}) likewise. 
In the translated coordinate system, the complete eigenvalue equation is given by

\begin{eqnarray}
&-&\frac{1}{2} \frac{d^2 u}{dx^2} + k [(1-\frac{4}{k}+\frac{3}{k^2})(1-2\frac{x}{\sqrt{k}}
+3\frac{x^2}{k} \nonumber \\
&-& 4\frac{x^3}{k^{3/2}}+...) + {r_0}^2 \{{\hat V}(r_0)+r_0 {\hat V}'(r_0)
(1+\frac{x}{\sqrt{k}}) \nonumber \\
&+& \frac{{r_0}^2}{2}{\hat V}''(r_0)(1+\frac{x^2}{k}+2\frac{x}{\sqrt{k}})
... \}] u(x) \nonumber \\
&=& (\frac{E}{k}) {r_0}^2 u(x)
\label{equation1}
\end{eqnarray}

In the analysis of the above equation we consider all terms up to $O(r^2)$ and evaluate
coefficients for increasing powers of $x$ starting with $x^0$. A little rearranging now allows
us to rewrite the eigenvalue equation in terms of the variable $x$ as follows

\begin{equation}
[H_0 + {\hat V}(x)] \psi(x) = \lambda \psi(x) 
\end{equation}

where

\begin{eqnarray}
H_0 &=& -\frac{1}{2} \frac{d^2}{dx^2} + \frac{1}{2} \omega^2 x^2 + \epsilon_0 \nonumber \\
{\hat V}(x) &=& \frac{1}{\sqrt{k}}(\epsilon_1 x + \epsilon_3 x^3) + \frac{1}{k}
(\epsilon_2 x^2 + \epsilon_4 x^4) \nonumber \\
&+& \frac{1}{k^{3/2}}(\delta_1 x + \delta_3 x^3
+ \delta_5 x^5) 
\label{potential}
\end{eqnarray}

where $\lambda=(\frac{E}{k}){r_0}^2$ and for a harmonic oscillator potential the 
constants $\epsilon_k$ and $\delta_k$ are given by 

\begin{eqnarray}
&& \epsilon_0 = \frac{k}{8}-\frac{1}{2}+\frac{3}{8 k}+\frac{k^2}{64} \nonumber \\
&& \epsilon_1=1,\:\:\epsilon_2=-3/2,\:\:\epsilon_3=\frac{1}{6}{r_0}^5 {\hat V}'''(r_0)-1/2
\end{eqnarray}

Higher-ordered parameters like $\epsilon_3$ have non-zero values for anharmonic oscillations.
The above description allows us to re-frame an effective classical potential 
$V_{\mathrm{eff}}$ in the large-N limit but now including higher-ordered fluctuations.
It has the form 

\begin{equation}
V_{\mathrm{eff}}(R) = -\frac{1}{2} \frac{\omega^2}{k} R^2 + \frac{\epsilon_0}{k} + 
\frac{1}{k} V(R)
\end{equation}

where $V(R)$ represents some oscillator potential having a minimum at $R_0$, a point which
can be obtained from the relation 

\begin{equation}
\frac{\partial}{\partial R} 
{V_{\mathrm{eff}}(R)}|_{R=R_0}=0
\label{optimization}
\end{equation}

Defining the potential as in eq. (\ref{potential}) and then applying the optimization criterion
as in eq. (\ref{optimization}), we arrive at the quadratic equation

\begin{equation}
3\epsilon_3 {R_0}^2 - \sqrt{k} \omega^2 R_0 + \epsilon_1 = 0
\label{quadratic}
\end{equation}

which gives the solution ${R_0}^{(\pm)}=\frac{\sqrt{k} \omega^2 \pm \sqrt{k \omega^4-12 \epsilon_1
\epsilon_3}}{6\epsilon_3}$. To check the stability at the point $R=R_0$, we evaluate
the second derivatives and find that the two roots of eq. (\ref{quadratic}) 
satisfy the relation

\begin{eqnarray}
&& \frac{\partial^2}{\partial R^2}{V_{\mathrm{eff}}(R)}|_{R={R_0}^{(\pm)}} \nonumber \\
&=& -\frac{\omega^2}{k}
+ \frac{6\epsilon_3}{k^{3/2}}(\frac{\sqrt{k} \omega^2 \pm 
\sqrt{k \omega^4-12 \epsilon_1 \epsilon_3}}{6\epsilon_3}) 
\label{second_derivatives}
\end{eqnarray}

The above result implies that the minima are subject to the restriction $k \omega^4 \geq 
12\epsilon_1 \epsilon_3$. An idea of the exactitude of this analysis can be had from an evaluation
of the parameters using a simple harmonic oscillator potential. This gives $\epsilon_3=-1/2$, 
thereby naturally validating the restriction. The conclusion remains unchanged even after 
adding higher ordered anharmonic terms to the potential. To leading order 
in expansions, we now have the large-N expanded energy eigenvalue for the ground state as
follows

\begin{equation}
E = \frac{k}{2} \frac{\omega^2}{{r_0}^2} {{R_0}^{(+)}}^2 + \frac{\sqrt{k}}{{r_0}^2} {R_0}^{(+)} 
(\epsilon_1 + \epsilon_3 {{R_0}^{(+)}}^2) + \frac{\epsilon_0}{k}
\label{energy}
\end{equation}

where ${r_0}^2=\frac{k}{2\omega}$. The above expression for energy conclusively proves that
even in the presence of fluctuations, large-N expansion gives positive corrections to energy,
monotonically approaching the exact value as one scales up the order. We have 
checked for a range of such higher ordered fluctuations and have found the previous conclusion
sacrosanct. A point of some interest here would be the variation of such an 
approximated energy with respect to the strength $\omega$ of the anharmonic oscillation
for a fixed dimension, N=3 say. Fig. 1 shows this variation and evidently tells us
that there is a minimum in the curve much as we would expect it to be. The minimum also
signifies the fact that the results of the large-N approximation would be best
valid close to the minimum, that is between $\omega=0.4-0.5$ as per Fig. 1.

\begin{figure}
\includegraphics[width=8.0cm, height=8.0cm, angle=0]{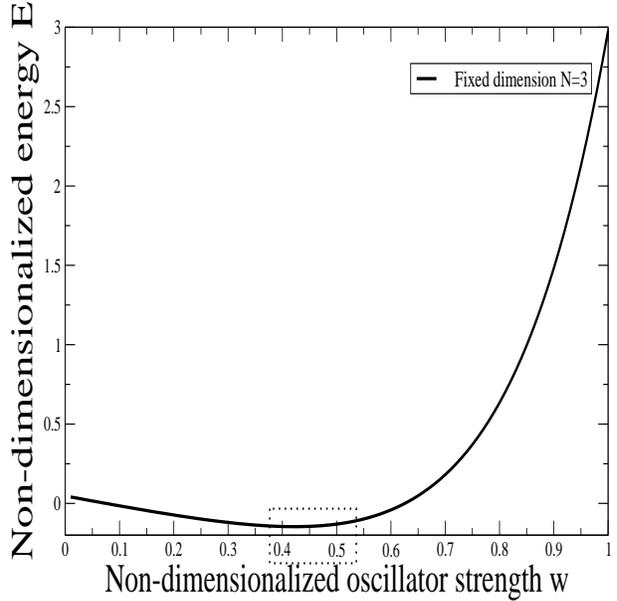}
\caption{Variation of non-dimensionalized energy E as in eq.(\ref{energy}) with 
the oscillator strength $\omega$ for $N=3$. The dotted line in the figure shows the
minimum around which the large-N approximation gives the best result.}
\label{fig1}
\end{figure}
 
As a suggestive example, we might look at the next higher modification in the
potential which gives rise to the following cubic equation

\begin{equation}
4\epsilon_4 {R_0}^3 + 3 \sqrt{k} \epsilon_3 {R_0}^2 + (2\epsilon_2-k\omega^2) R_0 
+\sqrt{k} \epsilon_1=0
\label{cubic}
\end{equation}

Once again, the above equation can be solved analytically using Cardan's method and it is rather
an easy algebraic exercise to show that the energy corrections are still positive. 

To conclude, we have shown using a perturbed anharmonic oscillator potential that a large-N 
expansion method provides an effective approximation scheme in tackling quantum mechanical
problems. This is evident, since the order of corrections as suggested by this method 
monotonically converges towards the semi-classical limit as $N \rightarrow \infty$. The 
results offer favorable comparisons with numerical and experimental data and might be
used in more complicated quantum many body problems \cite{chattopadhyay} involving
exact interaction potentials.

The author acknowledges helpful discussions with A. Chatterjee and is grateful to the 
Marie Curie Foundation, fellowship MIFI-CT-2005-008608, for research support.

\end{document}